\documentclass{article}
\title{Semiclassical description of relativistic spin without use of Grassmann variables and the Dirac equation}
\author{A. A. Deriglazov\footnote{alexei.deriglazov@ufjf.edu.br ~ On leave of
absence from Dept. Math. Phys., Tomsk Polytechnical University,
Tomsk, Russia.}}
\date{Dept. de Matematica, ICE, Universidade Federal de Juiz de Fora,
MG, Brazil.} 

\begin{document}
\maketitle
\large

\begin{abstract}
We propose a relativistic particle model without Grassmann variables which, being canonically quantized, leads to the
Dirac equation. Both $\Gamma$\,-matrices and the relativistic spin tensor are produced through the canonical
quantization of the classical variables which parametrize the properly constructed relativistic spin surface. Although
there is no mass-shell constraint in our model, our particle's speed cannot exceed the speed of light. The classical
dynamics of the model is in correspondence with the dynamics of mean values of the corresponding operators in the Dirac
theory. In particular, the position variable experiences {\it Zitterbewegung} in noninteracting theory. The classical
equations for the spin tensor are the same as those of the Barut-Zanghi model of a spinning particle.
\end{abstract}

\maketitle
\noindent
\section{Introduction}
Starting from the classical works [1-4], a lot of effort has been
spent on attempts to understand the behavior of a particle with
spin on the basis of semiclassical mechanical models [5-16,
21-27]. One possibility is to use the Grassmann (anticommutative)
variables for parametrization of the spin space [11, 12]. In their
pioneer work [11], Berezin and Marinov have suggested such a kind
model. Their prescription can be briefly summarized as follows.
For nonrelativistic spin, the noninteracting Lagrangian reads
$\frac{m}{2}(\dot x_i)^2+\frac{i}{2}\xi_i\dot\xi_i$, where the
spin inner space is constructed from vector-like Grassmann
variables $\xi_i$, $\xi_i\xi_j=-\xi_j\xi_i$. Since the Lagrangian
is linear on $\dot\xi_i$, their conjugate momenta coincide with
$\xi$, $\pi_i=\frac{\partial L}{\partial\dot\xi_i}=i\xi_i$. The
relations represent the Dirac second-class constraints and are
taken into account by the transition from the Poisson (Grassmann)
bracket to the Dirac one, the latter reading
\begin{eqnarray}\label{00}
\{\xi_i, \xi_j\}_{DB}=i\delta_{ij}.
\end{eqnarray}
Dealing with the Dirac bracket, one can resolve the constraints,
excluding the momenta from consideration. This gives a very
economic scheme for the description of a spin: there are only
three spin variables $\xi_i$ with the desired brackets (\ref{00}).
Canonical quantization is performed, replacing the variables by
the spin operators proportional to the Pauli $\sigma$-matrices,
$\hat S_i=\frac{\hbar}{2}\sigma^i$ ($[\sigma^i,
\sigma^j]_{+}=2\delta^{ij}$):
\begin{eqnarray}\label{01}
[\hat S_i, \hat S_j]_{+}=\frac{\hbar^2}{2}\delta_{ij}.
\end{eqnarray}
They act on the two-dimensional spinor space $\Psi_\alpha$. Canonical quantization of the particle on an external
electromagnetic background leads to the Pauli equation
\begin{eqnarray}\label{02}
i\hbar\frac{\partial\Psi}{\partial t}=\left(\frac{1}{2m}(\hat{{\bf
p}}-\frac{e}{c}{\bf A})^2-eA_0-
\frac{e\hbar}{2mc}\mbox{\boldmath$\sigma$}{\bf B}\right)\Psi.
\end{eqnarray}
Relativistic spin is described in a similar way [11, 12].

The problem here is that the Grassmann classical mechanics represents a rather formal mathematical construction. It
leads to certain difficulties [11, 16] in attempts to use it for describing the spin effects on the semiclassical
level, before the quantization. Hence it would be interesting to describe spin on the basis of usual variables. While
the problem has a long history (see [5-13] and references therein), there appears to be no wholly satisfactory solution
to date. It seems to be surprisingly difficult [13] to construct, in a systematic way, a consistent model that would
lead to the Dirac equation in the the course of canonical quantization. It is the aim of this work to construct an
example of a mechanical model for the one-particle sector of the Dirac equation.

To describe the nonrelativistic spin using commuting variables, we need to realize the commutator algebra of the
operators $\hat S_i$
\begin{eqnarray}\label{03}
[\hat S_i, \hat S_j]_{-}=i\hbar\epsilon_{ijk}\hat S_k,
\end{eqnarray}
instead of the anticommutator one (\ref{01}). This has been achieved in the recent work [15] starting from the
Lagrangian
\begin{eqnarray}\label{04}
S=\int dt\left(\frac{m}{2}(\dot x_i)^2+\frac{e}{c}A_i\dot x_i+eA_0
\right. \qquad \qquad \cr
+\left.\frac{1}{2g}(\dot\omega_i-\frac{e}{mc}\epsilon
_{ijk}\omega_jB_k)^2+\frac{3g\hbar^2}{8a^2}+
\frac{1}{\phi}((\omega_i)^2-a^2)\right).
\end{eqnarray}
The configuration-space variables are $x_i(t)$, $\omega_i(t)$, $g(t)$ and $\phi(t)$. Here $x_i$ represents the spatial
coordinates of the particle with the mass $m$ and the charge $e$, $\omega_i$ are the spin-space coordinates, $g$,
$\phi$ are the auxiliary variables, $a=\mbox{const}$ and ${\bf B}= \mbox{\boldmath$\nabla$}\times{\bf A}$. The second
and third terms in Eq. (\ref{04}) represent minimal interaction with the vector potential $A_0$, $A_i$ of an external
electromagnetic field, while the fourth term contains the interaction of the spin with a magnetic field. At the end, it
produces the Pauli term in the quantum mechanical Hamiltonian.

The Dirac constraints presented in the model imply [15] that spin lives on a two-dimensional surface of a
six-dimensional spin phase space $\omega_i$, $\pi_i$. The surface can be parametrized with the angular-momentum
coordinates $S_i=\epsilon_{ijk}\omega_j\pi_k$, subject to $S^2=\frac{3\hbar^2}{4}$.  They obey the classical brackets
$\{S_ i, S_j\}=\epsilon_{ijk}S_k$. Hence we quantize them according to the rule $S_i\rightarrow\hat S_i$.

The action leads to a reasonable picture on both classical and quantum levels. The classical dynamics is governed by
the Lagrangian equations
\begin{eqnarray}\label{05}
m\ddot x_i=eE_i+\frac{e}{c}\epsilon_{ijk}\dot
x_jB_k-\frac{e}{mc}S_k\partial_iB_k,
\end{eqnarray}
\begin{eqnarray}\label{06}
\dot S_i=\frac{e}{mc}\epsilon_{ijk}S_jB_k.
\end{eqnarray}
We have defined ${\bf E}=-\frac{1}{c}\frac{\partial{\bf A}}{\partial t}+\mbox{\boldmath$\nabla$}A_0$. Since
$S^2\approx\hbar^2$, the $S$-term disappears from Eq. (\ref{05}) in the classical limit $\hbar\rightarrow 0$. Then Eq.
(\ref{05}) reproduces the classical motion on an external electromagnetic field. Notice also that in the absence of
interaction, the spinning particle does not experience an undesirable {\it Zitterbewegung}. Equation (\ref{06})
describes the classical spin precession in an external magnetic field. On the other hand, canonical quantization of the
model immediately produces the Pauli equation (\ref{02}).

Below, we generalize this scheme to the relativistic case, taking angular-momentum variables as the basic coordinates
of the spin space. On this basis we construct the relativistically invariant classical mechanics that produces the
Dirac equation through the canonical quantization, and briefly discuss its classical dynamics.

\section{Relativistic spin surface}
Here we motivate our choice of the variables for describing the
relativistic spin. The dynamical model based on these variables is
constructed in the next section. The relativistic equation for the
spin precession can be obtained including the three-dimensional
spin vector $S_i$ (\ref{03}) either into the Frenkel tensor
$\Phi^{\mu\nu}$, $\Phi^{\mu\nu}u_\nu=0$, or into the
Bargmann-Michel-Telegdi $4$-vector $S^\mu$, $S^\mu u_\mu=0$, where
$u_\mu$ is the $4$-velocity vector\footnote{The conditions
$\Phi^{\mu\nu}u_\nu=0$ and $S^\mu u_\mu=0$ guarantee that in the
rest frame, only three components of these quantities survive,
which implies the right nonrelativistic limit.}. Unfortunately,
the semiclassical models based on these schemes do not lead to a
reasonable quantum theory, as they do not produce the Dirac
equation through the canonical quantization. We now motivate as to
how this can be achieved in the formulation that implies the
inclusion of $S_i$ into the $SO(2, 3)$ angular-momentum tensor
$J^{AB}$ of five-dimensional space $A=(\mu, 5)=(0, 1, 2, 3, 5)$,
$\eta^{AB}=(- + + + -)$.

In the passage from nonrelativistic to relativistic spin, we
replace the Pauli equation by the Dirac one:
\begin{eqnarray}\label{1.1}
(\hat p_\mu\Gamma^\mu+mc)\Psi(x^\mu)=0,
\end{eqnarray}
where $\hat p_\mu=-i\hbar\partial_\mu$. The position of the particle is described in the standard way; the
corresponding phase-space variables are $x^\mu$, $p^\nu$, $\{x^\mu, p^\nu\}_{PB}=\eta^{\mu\nu}$, $\eta^{\mu\nu}=(- + +
+)$.

Let us look for the classical variables that could produce the
$\Gamma$\,-matrices. According to the canonical quantization
paradigm, the classical variables, say $z^\alpha$, corresponding
to the Hermitian operators $\hat z^\alpha$ must be chosen to obey
the quantization rule
\begin{eqnarray}\label{0}
[\hat z^\alpha , \hat z^\beta]_{-}=i\hbar\left.\{z^\alpha , z^\beta \}\right|_{z\rightarrow\hat z}.
\end{eqnarray}
In this equation, $[ ~ , ~]_{-}$ is the commutator of the
operators and $\{ ~ , ~ \}$ stands for the classical
bracket\footnote{It is the Poisson (Dirac) bracket in a theory
without (with) second-class constraints.}. To avoid the
operator-ordering problems, we will consider only the sets of
operators which form the Lie algebra, $[\hat z^\alpha , \hat
z^\beta]_{-}=c^{\alpha\beta}{}_\gamma\hat z^\gamma$. So our first
task is to study the algebra of $\Gamma$\,-matrices. We note that
commutators of $\Gamma^\mu$ do not form closed Lie algebra, but
produce $SO(1, 3)$ Lorentz generators
\begin{eqnarray}\label{1.2}
[\Gamma^\mu, \Gamma^\nu]_{-}=-2i\Gamma^{\mu\nu},
\end{eqnarray}
where $\Gamma^{\mu\nu}\equiv\frac{i}{2}(\Gamma^\mu\Gamma^\nu-\Gamma^\nu\Gamma^\mu)$. The set $\Gamma^\mu$,
$\Gamma^{\mu\nu}$ forms a closed algebra. Besides the commutator (\ref{1.2}), we have
\begin{eqnarray}\label{1.3}
[\Gamma^{\mu\nu}, \Gamma^\alpha]_{-}=2i(\eta^{\mu\alpha}\Gamma^\nu-\eta^{\nu\alpha}\Gamma^\mu), \qquad \qquad \cr
[\Gamma^{\mu\nu}, \Gamma^{\alpha\beta}]_{-}=2i(\eta^{\mu\alpha}\Gamma^{\nu\beta}- \eta^{\mu\beta}\Gamma^{\nu\alpha}-
\eta^{\nu\alpha}\Gamma^{\mu\beta}+\eta^{\nu\beta}\Gamma^{\mu\alpha}).
\end{eqnarray}
The algebra can be identified with $SO(2, 3)$ Lorentz algebra with generators $\hat L^{AB}$:
\begin{eqnarray}\label{1.4}
[\hat L^{AB}, \hat L^{CD}]_{-}=2i(\eta^{AC}\hat L^{BD}-\eta^{AD}\hat L^{BC}-\eta^{BC}\hat L^{AD}+\eta^{BD}\hat L^{AC}),
\end{eqnarray}
assuming that $\Gamma^\mu\equiv\hat L^{5\mu}$, $\Gamma^{\mu\nu}\equiv\hat L^{\mu\nu}$.

According to Eqs. (\ref{0}) and (\ref{1.4}) we need classical variables with the algebra
\begin{eqnarray}\label{1.4.1}
\{J^{AB}, J^{CD}\}=2(\eta^{AC} J^{BD}-\eta^{AD}J^{BC}-\eta^{BC}J^{AD}+\eta^{BD}J^{AC}).
\end{eqnarray}
The problem is that in classical mechanics the basic phase-space
variables, say $\omega^\alpha, \pi_\beta$, necessarily obey the
Poisson bracket $\{\omega^\alpha,
\pi_\beta\}=\delta^\alpha{}_\beta$. The algebra differs from
(\ref{1.4.1}). So we generally need to pass from the initial to
some composed variables as well as to impose some constraints.
This implies the use of the Dirac machinery for constrained
theories [16, 17].

To arrive at the algebra (\ref{1.4.1}), we introduce, tentatively,
the ten-dimensional "phase" space of the spin degrees of freedom,
$\omega^A$, $\pi^B$, equipped with the Poisson bracket
\begin{eqnarray}\label{1.5}
\{\omega^A, \pi^B\}_{PB}=\eta^{AB},
\end{eqnarray}
and define the inner angular momentum
\begin{eqnarray}\label{1.6}
J^{AB}\equiv 2(\omega^A\pi^B-\omega^B\pi^A).
\end{eqnarray}
The Poisson brackets of these quantities form the desired classical algebra (\ref{1.4.1}).

Below we use the decompositions $(i, j=1, 2, 3)$
\begin{eqnarray}\label{1.61}
J^{AB}=(J^{5\mu}, J^{\mu\nu})=(~ J^{50},\, J^{5i}={\bf J}^5,\, J^{0i}={\bf W},\, J^{ij}=\epsilon^{ijk}D^k ~).
\end{eqnarray}

Since the $J^{AB}$ are the variables which we are interested in,
we try to take them as coordinates of the space $\omega^A, \pi^B$.
The Jacobian of the transformation $(\omega^A, \pi^B)\rightarrow
J^{AB}$ has rank equal to $7$. So, only seven among the ten
functions $J^{AB}(\omega, \pi)$, $A<B$, are independent
quantities. They can be separated as follows. By construction,
they obey the identity
\begin{eqnarray}\label{1.62}
\epsilon^{\mu\nu\alpha\beta}J^5{}_\nu J_{\alpha\beta}=0,
\Leftrightarrow
J^{ij}=(J^{50})^{-1}(J^{5i}J^{0j}-J^{5j}J^{0i}),
\end{eqnarray}
that is the $3$-vector $\bf{D}$ can be presented through ${\bf J}^5$, ${\bf W}$
\begin{eqnarray}\label{1.63}
{\bf{D}}=\frac{1}{J^{50}}{\bf{J}}^5\times{\bf W}.
\end{eqnarray}
Hence we can take $J^{5\mu}$ and $J^{0i}$ as a part of the new coordinate system. Suppose, we complete this set up to a
basis of the phase space adding three more coordinates, say $a, b, c$. Quantizing the complete set we obtain, besides
the desired operators $\hat L^{5\mu}, \hat L^{0i}$, some extra operators $\hat a, \hat b, \hat c$. They are not present
in the Dirac theory, and are not necessary for the description of the spin. So we need to reduce the dimension of our
space from $10$ to $7$, imposing three constraints. There is one important restriction on the choice of constraints.
Canonical quantization of a system with constraints implies replacement of the Poisson by the Dirac bracket; the latter
is constructed with the help of the constraints. We need $SO(2, 3)$\,-invariant constraints $T_a$, $\{T_a,
J^{AB}\}_{PB}=0$; otherwise the Dirac-bracket algebra will not coincide with those of the Poisson, (\ref{1.4.1}).

The only quadratic $SO(2, 3)$\,-invariants which can be
constructed from $\omega^A$, $\pi^B$ are $\omega^A\omega_A$,
$\omega^A\pi_A$ and $\pi^A\pi_A$. So we restrict our model to
living on the surface defined by the equations
\begin{eqnarray}\label{1.7}
T_3=\pi^A\pi_A+a_3=0
\end{eqnarray}
\begin{eqnarray}\label{1.71}
T_4=\omega^A\omega_A+a_4=0, \qquad T_5=\omega^A\pi_A=0,
\end{eqnarray}
where $a_3$, $a_4$ are some numbers. Our suggestion is to take the
surface as the inner space for the description of the relativistic
spin.

The matrix $\frac{\partial(J^{5\mu}, J^{0i}, T_4, T_5, \omega^5)}{\partial(\omega^A, \pi^B)}$ has rank equal ten. So
the quantities
\begin{eqnarray}\label{1.71.1}
J^{5\mu}, ~ J^{0i}, ~ T_4, ~ T_5, ~ \omega^5,
\end{eqnarray}
can be taken as coordinates of the space $(\omega^A$, $\pi^B)$.
The equation $J^{AB}=2(\omega^A\pi^B-\omega^B\pi^A)$ implies the
identity
\begin{eqnarray}\label{1.68}
J^{AB}J_{AB}=8[(\omega^A)^2(\pi^B)^2-(\omega^A\pi_A)^2]= \cr
8[(T_4-a_4)(T_3-a_3)-(T_5)^2], \quad
\end{eqnarray}
then the constraint $T_3$ can be written in the coordinates
(\ref{1.71.1}) as follows:
\begin{eqnarray}\label{1.71.2}
T_3=\frac{(J^{AB})^2+8(T_5)^2}{8(T_4-a_4)}+a_3,
\end{eqnarray}
where $J^{ij}$ are given by Eq. (\ref{1.61}). Note that $T_3$ does not depend on $\omega^5$. On the hyperplane
$T_4=T_5=0$ it reduces to
\begin{eqnarray}\label{1.71.3}
-8a_4T_3=(J^{AB})^2-8a_3a_4=0.
\end{eqnarray}
Eq. (\ref{1.71.3}) states that the value of $SO(2, 3)$\,-Casimir
operator\footnote{In quantum theory, for the operators
(\ref{1.4}), (\ref{1.3}) we have: $\hat J^{AB}\hat
J_{AB}=20\hbar^2$.} $(J^{AB})^2$ is equal to $8a_3a_4$.

In the dynamical model constructed below, the equation $T_3=0$ appears as the first-class constraint. It implies that
we are dealing with a theory with local symmetry, with the constraint being the generator of the symmetry [20]. The
coordinate $\omega^5$ is not inert under the symmetry, $\delta\omega^5\sim\{T_3, \omega^5\}=\ne 0$. Hence $\omega^5$ is
gauge non-invariant variable.

Summing up, we have restricted dynamics of spin on the surface (\ref{1.7}), (\ref{1.71}). If (\ref{1.71.1}) are taken
as coordinates of the phase space, the surface is the hyperplane $T_4=T_5=0$ with the coordinates $J^{5\mu}, J^{0i},
\omega^5$ subject to the condition (\ref{1.71.3}). Since $\omega^5$ is gauge non-invariant coordinate, we can discard
it.

This implies that we can quantize $J^{5\mu}$, $J^{0i}$ instead of the initial variables $\omega^A$, $\pi^B$. Similarly
to the case for the $\Gamma$\,-matrices, the brackets of the variables $J^{5\mu}$, $J^{0i}$ do not form a closed Lie
algebra. The nonclosed brackets are
\begin{eqnarray}\label{1.72.4}
\{J^{5i}, J^{5j}\}=\{J^{0i}, J^{0j}\}=-2J^{ij},
\end{eqnarray}
where $J^{ij}=(J^{50})^{-1}(J^{5i}J^{0j}-J^{5j}J^{0i})$; see Eq. (\ref{1.62}). Adding them to the initial variables, we
obtain the set $J^{AB}=(J^{5\mu}, J^{0i}, J^{ij})$ which obeys the desired algebra (\ref{1.4.1}). According to Eqs.
(\ref{0}), (\ref{1.4}), (\ref{1.4.1}), quantization is achieved by replacing $J^{AB}$ on the
$\Gamma$\,-matrices\footnote{The matrices $\Gamma^\mu$, $\Gamma^{\mu\nu}$ are Hermitian operators with respect to the
scalar product $(\Psi_1, \Psi_2)=\Psi_1^\dagger\Gamma^0\Psi_2$.}. We assume that $\omega^A$ has the dimension of
length; then $J^{AB}$ has the dimension of the Planck's constant. Hence the quantization rule is
\begin{eqnarray}\label{1.72}
J^{5\mu}\rightarrow\hbar\Gamma^\mu, \quad J^{\mu\nu}\rightarrow\hbar\Gamma^{\mu\nu}.
\end{eqnarray}

This implies, that the Dirac equation (\ref{1.1}) can be produced through the constraint
\begin{eqnarray}\label{1.8}
T_2\equiv p_\mu J^{5\mu}+mc\hbar=0.
\end{eqnarray}
Summing up, to describe the relativistic spin, we need a theory that implies the Dirac constraints (\ref{1.7}),
(\ref{1.71}), (\ref{1.8}) in the Hamiltonian formulation.

\section{Lagrangian action and the canonical quantization}
One possible dynamical realization of the construction presented above is given by the following $d=4$
Poincar\'e-invariant Lagrangian
\begin{eqnarray}\label{2.1}
L=-\frac{1}{2e_3} \left[(\dot
x^\mu+e_2\omega^\mu)^2-(e_2\omega^5)^2\right]-\frac{\sigma
mc\hbar}{2\omega^5}-\frac{\sigma^2a_3}{2e_3}+\cr
\frac{1}{\sigma}\left[(\dot
x^\mu+e_2\omega^\mu)\dot\omega_\mu-e_2\omega^5\dot\omega^5\right]-
e_4(\omega^A\omega_A+a_4),
\end{eqnarray}
written on the configuration space $x^\mu$, $\omega^\mu$,
$\omega^5$, $e_i$, $\sigma$, where $e_i$, $\sigma$ are the
auxiliary variables. The local symmetries of the theory are the
reparametrizations as well as the following transformations with
the parameter $\epsilon(\tau)$ (below we have defined
$\beta\equiv\dot e_4\epsilon+\frac12e_4\dot\epsilon$)
\begin{eqnarray}\label{2.2}
\delta x^\mu=0, \quad \delta\omega^A=\beta\omega^A, \quad
\delta\sigma=\beta\sigma, \quad \delta e_3=0, \cr \delta
e_2=-\beta e_3+\frac{e_2}{\sigma}\dot\beta, \quad \delta
e_4=-2e_4\beta-\left(\frac{e_2\dot\beta}{2\sigma^2}\right)\dot{}.
\end{eqnarray}
The presence of local symmetries implies the appearance of the
first-class constraints (\ref{1.7}), (\ref{1.8}) in the
Hamiltonian formalism. We point out that in the Berezin-Marinov
model the Dirac equation is implied by the supersymmetric gauge
transformations. So, the symmetry (\ref{2.2}) represents the
bosonic analogue of these transformations.

Although there is no the constraint $p^2+ m^2c^2=0$ in our model,
our particle's speed cannot exceed the speed of light. Indeed,
equations of motion for the auxiliary variables $e_2$, $e_3$ read
$(\dot x\omega)=e_2a_4$, $\dot x^2+2e_2(\dot
x\omega)+e_2^2\omega^A\omega_A+\sigma^2a_3=0$. They imply $(\dot
x^\mu)^2=-e_2^2a_4-\sigma^2a_3$. Since we are dealing with a
reparametrization-invariant theory, we assume that the functions
$x^\mu(\tau)$ represent the parametric equations of the trajectory
$x^i(t)$. Then the previous equation implies
\begin{eqnarray}\label{2.21}
\left(\frac{dx^i}{dt}\right)^2=\left(c\frac{\dot x^i(\tau)}{\dot
x^0(\tau)}\right)^2=c^2\left(1-\frac{e_2^2a_4+\sigma^2a_3}{(\dot
x^0)^2}\right)<c^2,
\end{eqnarray}
if we take $a_3>0$, $a_4>0$.

Curiously enough, the Lagrangian (\ref{2.1}) can be rewritten in
almost five-dimensional form. Namely, after the change
\footnote{The change is an example of conversion of the
second-class constraints in the Lagrangian formulation [18].}
$(x^\mu, \sigma, e_2)$ $\rightarrow$ $(\tilde x^\mu, \tilde x^5,
\tilde e_2)$, where $\tilde
x^\mu=x^\mu-\frac{e_3}{\sigma}\omega^\mu$, $\tilde
x^5=-\frac{e_3}{\sigma}\omega^5$, $\tilde
e_2=e_2+(\frac{e_3}{\sigma})\dot{}$, it reads
\begin{eqnarray}\label{2.1.1}
L=-\frac{1}{2e_3}(D\tilde x^A)^2+\frac{(\tilde
x^5)^2}{2e_3(\omega^5)^2}(\dot\omega^A)^2+\frac{e_3mc\hbar}{\tilde
x^5}+\frac{e_3a_3(\omega^5)^2}{2(\tilde x^5)^2}- \cr
e_4(\omega^A\omega_A+a_4), \qquad \qquad \qquad \qquad
\end{eqnarray}
where we have defined $D\tilde x^A=\dot{\tilde x}^A+\tilde e_2\omega^A$.

\par
\noindent {\it Canonical quantization.} To confirm that the action
(\ref{2.1}) leads to the Dirac equation, we construct its
Hamiltonian formulation. In the Hamiltonian formalism, the
equations (\ref{1.7}), (\ref{1.71}), (\ref{1.8}) appear as the
Hamiltonian constraints. The constraint $T_2$ has vanishing
Poisson brackets with all the constraints. The remaining
constraints obey the Poisson-bracket algebra
\begin{eqnarray}\label{Z2.1_02}
\{T_3, T_4\}=-4T_5, \qquad \{T_3, T_5\}=-2T_3+2a_3, \cr \{T_4,
T_5\}=2T_4-2a_4. \qquad \qquad \qquad
\end{eqnarray}
If we take the combination
\begin{eqnarray}\label{Z2.1_03}
\tilde T_3\equiv T_3+\frac{a_3}{a_4}T_4,
\end{eqnarray}
the algebra acquires the form
\begin{eqnarray}\label{Z2.1_01}
\{\tilde T_3, T_4\}=-4T_5, \quad \{\tilde T_3,
T_5\}=-2T_3+2\frac{a_3}{a_4}T_4, \cr \{T_4, T_5\}=2T_4-2a_4.
\qquad \qquad
\end{eqnarray}
The only bracket which does not vanish on the constraint surface
is $\{T_4, T_5\}$.  According the Dirac terminology [16, 17], we
have the first-class constraints (\ref{Z2.1_03}), (\ref{1.8}), and
the second-class pair (\ref{1.71}). The presence of the
first-class constraints indicates that we are dealing with a
theory invariant under a two-parameter group of local (gauge)
symmetries, which has been discussed above.

The auxiliary variables $\sigma$, $e_4$ turn out to be subject to
their own second-class constraints. Assuming that the constraints
are taken into account by the transition from the Poisson to the
Dirac bracket, the variables can be omitted from consideration
[16, 17]. The Hamiltonian in terms of the remaining variables
reads
\begin{eqnarray}\label{2.3}
H=\frac{1}{2e_3}\left(\frac{\omega^5e_2}{\pi^5}\right)^2(\pi^A\pi_A+a_3)+\frac{e_2}{2\pi^5}(p_\mu
J^{5\mu}+mc\hbar)+ \cr \lambda_{e2}\pi_{e2}+\lambda_{e3}\pi_{e3}.
\qquad \qquad \qquad \qquad
\end{eqnarray}
Here $\pi_{ea}$,  are conjugate momenta for $e_{a}$ and
$\lambda_{ea}$ are the Lagrangian multipliers for the primary
constraints $\pi_{ea}=0$.

The constraints (\ref{1.7}), (\ref{1.71}) can also be taken into
account by using of the Dirac bracket. Since they represent $SO(2,
3)$\,-invariants, the Dirac brackets of the quantities $J^{AB}$
coincide with the Poisson one, Eq. (\ref{1.4.1}). Hence we
quantize the model according to Eq. (\ref{1.72}). In the quantum
theory, the first-class constraint (\ref{1.8}) is imposed on the
state vector. This gives the Dirac equation. In the result,
canonical quantization of the model leads to the desired quantum
picture.

\section{Solution to the classical equations of motion}
We now discuss some properties of the classical theory and confirm
that they are in correspondence with those of the one-particle
sector of the Dirac equation [19, 20].

Besides the constraints discussed above, the Hamiltonian
(\ref{2.3}) implies the following equations (we use the notation
$(p\omega)=p^\mu\omega_\mu$\,):
\begin{eqnarray}\label{Z2.2}
\dot e_a=\lambda_{ea}, \quad \pi_{ea}=0, \quad a=2, 3;
\end{eqnarray}
\begin{eqnarray}\label{Z2.3}
\dot x^\mu=\frac{e_2}{2\pi^5}J^{5\mu}, \quad \dot p^\mu=0;
\end{eqnarray}
\begin{eqnarray}\label{Z2.4}
\dot\omega^\mu=\frac{1}{e_3}\left(\frac{e_2\omega^5}{\pi^5}\right)^2\pi^\mu+
\frac{e_2\omega^5}{\pi^5}p^\mu, \quad
\dot\pi^\mu=e_2p^\mu-2e_4\omega^\mu; \quad \cr
\dot\omega^5=\frac{1}{e_3}\left(\frac{e_2\omega^5}{\pi^5}\right)^2\pi^5+\frac{e_2}{\pi^5}(p\omega),
\quad \dot\pi^5=\frac{e_2}{\pi^5}(p\pi)-2e_4\omega^5,
\end{eqnarray}
Here
$e_4=\frac{1}{2e_3}\left(\frac{e_2\omega^5}{\pi^5}\right)^2\frac{a_3}{a_4}$.
The equations reflect the fact that we are dealing with a theory
with local symmetries. Indeed, we note that these equations do not
determine the Lagrangian multipliers $\lambda_{ea}$, which enter
as arbitrary functions into solutions to the equations of motion
for $x^\mu$, $\omega^A$, $\pi^A$. According to the general theory
[16, 17], variables with ambiguous dynamics do not represent the
observable quantities. For our case, all the variables except
$p_\mu$ turn out to be ambiguous.

To construct the gauge-invariant variables with unambiguous
dynamics we first note that $x^\mu$, $p_\mu$, $J^{5\mu}$ and
$J^{\mu\nu}$ represent $\epsilon$\,-invariant quantities.
Equations for the angular-momentum variables follow from
(\ref{Z2.4})
\begin{eqnarray}\label{2.4b}
\dot J^{5\mu}=-\frac{e_2}{\pi^5}J^{\mu\nu}p_\nu, \quad
\dot J^{\mu\nu}=\frac{e_2}{\pi^5}(p^\mu J^{5\nu}-p^\nu J^{5\mu}).
\end{eqnarray}
In three-dimensional notation, these equations read
\begin{eqnarray}\label{2.4f}
\dot J^{50}=-({\bf W}{\bf p}), \quad \dot{\bf J}^5=-p^0{\bf
W}+{\bf D}\times{\bf p}, \cr
\dot{\bf W}=p^0{\bf J}^5-J^{50}{\bf p}. \qquad \qquad
\end{eqnarray}
The ambiguity remaining in Eqs. (\ref{Z2.3}), (\ref{2.4b}) due to
the factor $\frac{e_2}{\pi^5}$ has a well-known interpretation,
being related to the reparametrization invariance of the theory.
We assume that, in accordance with this, the functions
$x^\mu(\tau)$, $J^{AB}(\tau)$ represent the dynamical variables
$x^i(t)$, $J^{AB}(t)$ in the parametric form. Using the identity
$\frac{dA(t)}{dt}=c\frac{\dot A(\tau)}{\dot x^0(\tau)}$, we obtain
the deterministic evolution for the gauge-invariant variables:
\begin{eqnarray}\label{2.8}
\frac{dx^i}{dt}=c(J^{50})^{-1}J^{5i}, \quad
\frac{dJ^{5\mu}}{dt}=-2c(J^{50})^{-1}J^{\mu\nu}p_\nu, \cr
\frac{dJ^{\mu\nu}}{dt}=2c(J^{50})^{-1}(p^\mu J^{5\nu}-p^\nu J^{5\mu}). \qquad \qquad
\end{eqnarray}

To find the trajectory $x^i(t)$, we take the equations
(\ref{Z2.3}), (\ref{2.4b}) in the gauge $\frac{e_2}{\pi^5}=1$ and
note that they imply the following closed third-order equation for
$x^\mu$
\begin{eqnarray}\label{2.9}
\stackrel{...}{x}^\mu-p^2\dot x^\mu=\frac{mc\hbar}{2}p^\mu.
\end{eqnarray}
For any $p^2<0$ its solution is given by
\begin{eqnarray}\label{2.10}
x^\mu(\tau)=x^\mu_0-\frac{mc\hbar}{2p^2}p^\mu\tau+\frac{a^\mu}{\sqrt{-p^2}}\cos(\sqrt{-p^2}\tau+\phi^\mu).
\end{eqnarray}
Since $x^0(\tau)$ in a reparametrization-invariant theory must be
a monotonic function of $\tau$, we take the integration constant
$a^0=0$. Then the parametric equations (\ref{2.10}) imply
\begin{eqnarray}\label{2.11}
x^i(t)=x^i_0+\frac{cp^i}{p_0}t+\frac{a^i}{\sqrt{-p^2}}\cos(\omega t+\phi^i).
\end{eqnarray}
The solution is a combination of the rectilinear motion and oscillations with the frequency
\begin{eqnarray}\label{2.12}
\omega=\frac{2\left(-p^2\right)^{\frac{3}{2}}}{m\hbar p^0}.
\end{eqnarray}
For the particular value $p^2=-m^2c^2$, and when $p^i\ll p^0$, it
approaches to the Compton frequency $\frac{2mc^2}{\hbar}$. Hence
the variable $x^i(t)$  experiences the {\it Zitterbewegung} in
noninteracting theory.

As we have discussed above, our particle's speed  cannot exceed
the speed of light for any value of $p^2$. For the case $p^2>0$,
the general solution to Eq. (\ref{2.9}) represents hyperbolic
motion
\begin{eqnarray}\label{2.13}
x^\mu(\tau)=x^\mu_0-\frac{mc\hbar}{2p^2}p^\mu\tau+\frac{a^\mu}{\sqrt{p^2}}\cosh(\sqrt{p^2}\tau+\phi^\mu).
\end{eqnarray}
The existence of such a self-accelerated solution for the Frenkel
electron has been recently observed; see [21].

\par
\noindent {\it The variables free of Zitterbewegung.} Besides the
centre of charge, $\hat x$, in the Dirac theory we can construct
the centre-of-mass (Pryce-Newton-Wigner) [6, 7] operator
$\hat{\tilde x}$ in such a way that the conjugated momentum of
$\hat x$ turns out to be the mechanical momentum for $\hat{\tilde
x}$. So the Dirac particle looks like a kind of composed system
(this picture has been used by Schr\"odinger [1] to identify spin
with inner angular momentum of the system). The classical analogue
of the centre-of-mass operator in our model is the variable
\begin{eqnarray}\label{2.13}
\tilde x^\mu=x^\mu+\frac{1}{2p^2}J^{\mu\nu}p_\nu.
\end{eqnarray}
It obeys the equation $\dot{\tilde x}^\mu(\tau)=-\frac{e_2mc\hbar}{2\pi^5p^2}p^\mu$; then the centre of mass $\tilde
x^i(t)$ moves along the straight line, $\frac{d\tilde x^i}{dt}=\frac{cp^i}{p^0}$.  Note also that $p^\mu$ represents
the mechanical momentum of the $\tilde x^\mu$\,-particle.

Let us take as the classical four-dimensional spin vector the
Pauli-Lubanski vector $S^\mu=\epsilon^{\mu\nu\alpha\beta}p_\nu
J_{\alpha\beta}$. It has no precession in the free theory; $\dot
S^\mu=0$. In the centre-of-charge instantaneous rest frame,
\begin{eqnarray}\label{2.5}
J^{50}=\mbox{const}, \quad {\bf J}^5=0,
\end{eqnarray}
it reduces to $S^0=0$, ${\bf S}={\bf p}\times{\bf W}$. According
to Eqs. (\ref{2.5}), (\ref{1.63}), only the part ${\bf W}$ of the
angular-momentum tensor (\ref{1.61}) survives in the
nonrelativistic limit.

\par
\noindent {\it Comparison with the Barut-Zanghi (BZ) model}. The BZ spinning particle [14] is widely used [22-27] for
semiclassical analysis of spin effects. Starting from the even variables $z_\alpha$, where $\alpha=1, 2, 3, 4$ is the
$SO(1, 3)$ spinor index, Barut and Zanghi have constructed an even spin tensor $S_{\mu\nu}=\frac{1}{4}i\bar
z\gamma_{\mu\nu}z$. We point out that in the gauge $\frac{e_2}{\pi^5}=1$ our equations (\ref{Z2.3}),(\ref{2.4b})
coincide with those of the BZ model, making the identifications $J^{5\mu}\leftrightarrow v^\mu$,
$J^{\mu\nu}\leftrightarrow S^{\mu\nu}$. Besides, our model implies the equations $(\frac{J^{5\mu}}{\pi^5})^2=-4R$,
$p_\mu J^{5\mu}+mc\hbar=0$. The first equation guarantees that the centre of charge cannot exceed the speed of light.
The second equation implies the Dirac equation\footnote{The Barut-Zanghi model [10] does not imply the Dirac
equation.}.

\section{Conclusion}

In this work we have constructed a semiclassical model (\ref{2.1}), (\ref{2.3}) which describes the one-particle sector
of the Dirac equation. Although there is no the constraint $p^2+ m^2c^2=0$ in our model, our particle's speed cannot
exceed the speed of light. Spinning degrees of freedom are described on the basis of a seven-dimensional surface
embedded in the ten-dimensional phase space $\omega^A$, $\pi^A$ equipped with the Poisson bracket (\ref{1.5}). The
surface is specified by the $SO(2, 3)$\,-invariant equations (\ref{1.7}), (\ref{1.71}). The angular-momentum variables
$J^{5\mu}$, $J^{0i}$ can be taken as coordinates of the surface. Quantizing them in accordance with their
Poisson-bracket algebra (\ref{1.4.1}), we have produced both the $\Gamma^\mu$\,-matrices and the relativistic spin
tensor $J^{\mu\nu}$. The first-class constraint (\ref{1.8}) is imposed on the state vectors, which leads to the Dirac
equation.

Our model shows the same undesirable properties as those of the Dirac equation in the semiclassical limit. We have
solved the classical equations of motion and confirmed that the position variable $x^i$ experiences the {\it
Zitterbewegung} in noninteracting theory, see (\ref{2.11}). The variable $\tilde x^\mu$ specified by the equation
(\ref{2.13}) moves along a straight line and corresponds to the Pryce-Newton-Wigner operator of the Dirac theory. Like
the Dirac equation, the model presented here gives no evidence as regards which of these two variables should be
identified with the particle position.

We finish with a brief comment on a modification which solves the problems. We recall that the Dirac equation
(\ref{1.1}) implies the Klein-Gordon one. In contrast, in classical mechanics the corresponding constraint (\ref{1.8})
does not imply the mass-shell constraint $p^2+m^2c^2=0$. So, the model presented here is not yet in complete
correspondence with the Dirac theory. The semiclassical model that produces both constraints has been discussed in the
recent work [28]. The extra first-class constraint implies that we are dealing with a theory with one more local
symmetry, with the constraint being a generator of the symmetry [29, 17]. This leads to a completely different picture
of the classical dynamics. The variable $x^\mu$ is not inert under the extra symmetry; $\delta x^\mu=\beta p^\mu$,
where $\beta(\tau)$ is the local parameter. Being gauge non-invariant, $x^\mu$ turns out to be an unobservable
quantity. The variable $\tilde x^\mu$ of Eq. (\ref{2.13}) is gauge invariant and should be identified with the position
of the particle. Because $p^\mu$ is a mechanical momentum for $\tilde x^\mu$, the particle's speed cannot exceed the
speed of light. In the absence of interaction it moves along the straight line. Hence the modified model is free of the
undesirable {\it Zitterbewegung}.

\section{Acknowledgments}
This work was supported by the Brazilian foundation FAPEMIG.

\end{document}